\documentclass[letterpaper,11pt,reqno]{amsart}

\makeatletter
\usepackage{amssymb}
\usepackage{latexsym}
\usepackage{amsbsy}
\usepackage{amsfonts}
\usepackage{xurl}
\usepackage{hyperref}
\usepackage{graphicx}
\usepackage{enumerate}
\usepackage{enumitem}
\usepackage{mathtools}
\usepackage{color}
\usepackage{booktabs}
\usepackage{tabularx}
\usepackage{array}

\usepackage{tikz}

\def\marginpar#1{\ignorespaces}

\DeclareMathOperator\var{Var}
\DeclareMathOperator\argmax{\arg \max}

\newtheorem{theorem}{Theorem}[section]

\newtheorem{proposition}[theorem]{Proposition}

\numberwithin{equation}{section}
\makeatother
\begin{document}
\title[RWA]{A prelude to the theory of Real-World Asset (RWA) Tokenization}

\author[Wenpin Tang]{{Wenpin} Tang}
\address{Department of Industrial Engineering and Operations Research, Columbia University. 
} \email{wt2319@columbia.edu}

\date{\today} 
\begin{abstract}
We provide an introduction to real-world asset (RWA) tokenization,
and develop two economic models connecting this emerging market to classical financial theory.
In search for a theory of RWA tokenization,
we classify RWA tokenization according to the role of the underlying asset, 
distinguishing standard RWA tokenization, stablecoins, and Digital Asset Treasuries (DATs).
Our first model connects tokenization with security design under asymmetric information.
Continuous secondary-market trading generates information recovery, 
reducing adverse-selection costs over time and affecting both the issuance price and the optimal tokenization design.
Our second model develops a dynamic corporate-finance framework for DATs, 
in which market valuation, capital-market financing, and crypto accumulation interact endogenously.
We characterize the optimal issuance policy and illustrate the model using \texttt{Strategy} as a case study.
A few directions for future research are discussed, providing a starting point for further theoretical and empirical study of RWA markets.
 \end{abstract}

\maketitle

%\setcounter{tocdepth}{1}
%\tableofcontents
%-------------------------------------------------------------------------------------------------
\section{Introduction}
\label{sc1}
\quad Over the past decade, blockchain technology has developed rapidly,
 with applications including 
cryptocurrencies \cite{Naka08, Wood14},
healthcare \cite{MC19, TPE20},
supply chain \cite{CTT20, Kam18},
and decentralized finance (DeFi) \cite{Uni2, Hyperliquid}.
One important consequence of this development is the revival of real-world asset (RWA) tokenization, 
which converts rights to an off-chain asset into digital tokens 
that can be issued, transferred, and traded electronically. 
Precursors to this idea date back to pre-blockchain digital-asset systems such as eGold\footnote{See \url{https://en.wikipedia.org/wiki/E-gold} for the history of eGold.}, 
but blockchain technology has substantially expanded its scope by providing 
compliance token standards, programmable transferability, and on-chain settlement.
More recently, token standards such as ERC-1400\footnote{See \url{https://github.com/ethereum/EIPs/issues/1411} for details on ERC-1400 for security token standards.} and ERC-3643\footnote{See \url{https://www.erc3643.org/} for details on ERC-3643 for RWA tokenization.}
have incorporated compliance requirements directly into token infrastructure.
Tokenization can therefore facilitate fractional ownership,
and make claims on traditional assets transferable through blockchain-based markets. 
See \cite{Bu23} for a brief history and development of the RWA industry.

\quad The economic scale of RWA tokenization can be substantial:
a Boston Consulting Group-ADDX report\footnote{See \url{https://addx.co/files/bcg_ADDX_report_Asset_tokenization_trillion_opportunity_by_2030_de2aaa41a4.pdf} for the report.} estimates that tokenized asset could reach $\$16.1$ trillion by 2030, 
or approximately $10\%$ of global GDP.
Existing RWA markets already cover a wide range of assets. 
For instance, \cite{Li25} classifies $180$ RWA products into six major categories:
private credit, stocks, global bonds, commodities, institutional funds, and U.S. Treasuries.
The composition of the RWA market has evolved rapidly.
While private credit was the dominant segment in the early development of RWA tokenization\footnote{In August 2025, the tokenized private credit was about $\$15.3$ billion of the $\$26.3$ billion non-stablecoin RWA market, compared with $\$7.5$ billion in tokenized U.S. Treasuries.},
tokenized U.S. Treasuries had become the largest category by August 2026, with approximately $\$16$ billion of the $\$38.1$ billion non-stablecoin RWA market, compared with about $\$7$ billion in tokenized private credit\footnote{The RWA data is available at \url{https://app.rwa-xyz.com/}.}. 
The scope of tokenization is also expanding toward liquid securities.
In January 2026, U.S. Securities and Exchange Commission issued a statement\footnote{See \url{https://www.sec.gov/newsroom/speeches-statements/corp-fin-statement-tokenized-securities-012826-statement-tokenized-securities} for SEC's statement on tokenized securities.} clarifying several structures for tokenized securities, including issuer-sponsored, custodial, and synthetic 
models.
In August 2026, Coinbase introduced tokenized versions of Apple, Nvidia, Meta, and Alphabet shares on its Base network\footnote{See \url{https://www.coindesk.com/business/2026/08/24/coinbase-debuts-tokenized-stocks-on-base-network-joining-race-to-bring-equities-on-blockchain}.}, while South Korea announced a roadmap toward a broader tokenized securities market beginning in 2027\footnote{See \url{https://www.coindesk.com/business/2026/09/04/south-korea-targets-february-2027-rollout-for-full-tokenized-securities-market}.}.
These developments suggest that tokenization evolve from a specialized application for illiquid assets toward a potentially broader infrastructure for financial markets. 
Refer to \cite{Cong25, Li25, Mir25, VPB26} for recent studies of RWA tokenization
on the legal and technological architectures and their policy implications.
We will also provide a brief review of this landscape in Section \ref{sc2}.

\quad The purpose of this paper is twofold.
First, we provide a self-contained taxonomy of RWA tokenization,
which differs from classifications based directly on asset classes.
Second, we develop simple economic models that isolate some of the key mechanisms underlying RWA tokenization,
connecting these new financial structures to classical security design, market liquidity and corporate finance.
Here, we propose to categorize tokenization by the economic role played by the RWA:
\begin{itemize}[itemsep = 3 pt]
\item
In the first category, the RWA is the {\bf object being tokenized}: stocks, bonds, private credit, real estate, commodities, and other assets are transformed into on-chain claims in order to facilitate ownership, financing, and trading. 
\item
In the second category, the RWA is the {\bf backing asset}, while the token itself is the object of interest. 
Stablecoins are the main example: Treasury securities or other reserve assets support a digital monetary liability designed to trade near par. 
\item
Finally, Digital Asset Treasuries (DATs) also provide a tokenization perspective. 
A DAT holds cryptoassets on its corporate balance sheet, and transforms their economic value into publicly traded corporate securities.
Though DATs do not fall within the conventional definition of RWA tokenization, 
they exhibit a closely related transformation between underlying assets and continuously traded market values.\footnote{As an example, \texttt{Strategy} is a DAT whose principal treasury asset is Bitcoin.
It fits both of the preceding categories.
From the investor's perspective, its publicly traded equity is a proxy for Bitcoin exposure, 
and hence loosely as a ``tokenized" claim on Bitcoin.
From the company's perspective, Bitcoin is the underlying treasury asset that supports its publicly traded corporate securities.
So Bitcoin is simultaneously the object of investors' exposure and an asset backing the corporation's financial claims.}
\end{itemize}

\quad There has been a substantial literation on stablecoins, 
including their reserve design, redemption, liquidity management and regulation,
which we briefly review in Section \ref{sc2}. 
In this paper, we focus on the other two directions:
\begin{enumerate}[itemsep = 3 pt]
\item
Our first model connects RWA tokenization with classical securitization and security design,
following the DeMarzo-Duffie framework \cite{DD99}.
A new feature that is particularly natural for tokenized assets is continuous secondary-market trading. 
Whereas traditional securitization is largely a security-design problem at issuance,
tokenization adds a dynamic trading layer through which order flow reveals information about the underlying asset. 
In our model, trading generates information recovery, 
and reduces the adverse-selection cost over time, 
which feeds back into the issuance price and optimal tokenization design. 
We characterize the resulting optimal tokenization design explicitly,
and discuss how it depends on information precision, market liquidity, and the issuer's financing opportunities.
\item
Our second model develops a dynamic corporate-finance framework for DATs, 
focusing on the interaction between market valuation, capital-market financing, and crypto accumulation.
Unlike a passive crypto ETF, a DAT can issue securities to acquire additional cryptoassets,
while its ability and incentive to do so depend on its market valuation relative to the net asset value (NAV) of its crypto holdings.
We show that a premium to NAV makes common-equity issuance accretive,
and derive an equilibrium in which the NAV premium, optimal issuance, and crypto accumulation are jointly determined.
We also extend the framework to preferred equity, debt, and convertible securities, 
and use \texttt{Strategy} as a case study to calibrate the model with market data.
\end{enumerate}

\quad This paper is both introductory and research-oriented.
On the one hand, we aim to provide a self-contained introduction to the rapidly evolving landscape of RWA tokenization,
 and its connections to traditional financial instruments.
On the other hand, 
we use this perspective as a starting point to develop simple models that capture some of its key mechanisms,
and identify problems that can be studied using market data. 
Instead of providing an exhaustive survey or a fully developed theory,
we aim to bridge emerging practices in RWA markets with classical ideas in financial economics.
We hope that this initial exploration will trigger further theoretical and empirical research on RWA tokenization.
Some of the introductory material in this paper is based on lectures by the author for IEOR E4704: Introduction to Financial Technology, at Columbia University.

\medskip
{\bf Organization}: The remainder of the paper is organized as follows.
Section \ref{sc2} provides background on RWA tokenization.
Section \ref{sc5} develops the tokenization model, while Section \ref{sc6} presents the DAT model.
Concluding remarks are given in Section \ref{sc4}.

%-------------------------------------------------------------------------------------------------
\section{Background on RWA tokenization}
\label{sc2}

\quad This section provides background on RWA tokenization, organized around the three categories introduced above: standard RWA tokenization, stablecoins, and DATs.

\subsection{RWA tokenization}
\label{sc21}

Standard RWA tokenization converts an off-chain asset to a digital token,
which gives its holder an economic claim on the underlying asset or its cash flows.
The underlying asset is the object of interest, 
while blockchain technology provides a new infrastructure for ownership, transfer, settlement, and potentially secondary market trading.
As emphasized by \cite{VPB26}, 
current RWA systems are hybrid rather than purely on-chain:
transfer and settlement may occur through blockchain infrastructure, 
while custody, compliance and verification continue to rely on off-chain institutions.

\quad The range of assets that can be tokenized is broad. 
\begin{itemize}[itemsep = 3 pt]
\item
Real estate is a natural example of tokenization, as fractional ownership can reduce the barriers associated with high asset values and illiquidity. 
For instance, \texttt{RealT} tokenizes interests in individual properties, providing investors with fractional ownership and rental income.
Its tokens can be transferred on blockchain-based secondary markets (e.g., \texttt{YAM}\footnote{See \url{https://yam.realtoken.network/} for the marketplace trading RealT tokens.}) subject to investor whitelisting and other compliance restrictions.
Commodities provide another example: \texttt{PAX Gold} (\texttt{PAXG})\footnote{See \url{https://docs.paxos.com/guides/stablecoin/paxg}.} represents physical gold held in custody, with each token corresponding to one troy ounce of gold.
\item
A second important category is private credit and other debt claims, 
where loans and other private debt can be represented by transferable blockchain claims.
As observed in \cite{Li25}, this category experiences rapid growth in the early RWA market.
However, significant frictions remain, particularly limited secondary-market trading,
because private credit is relatively illiquid and informationally opaque.
\item
A third category is institutional tokenized financial assets, especially U.S. Treasuries and money-market funds. 
Examples include Ondo's \texttt{OUSG}\footnote{See \url{https://docs.ondo.finance/qualified-access-products/ousg/}.}, which provides tokenized exposure to U.S. Treasuries; 
BlackRock's \texttt{BUIDL}\footnote{See \url{https://investors.securitize.io/news/news-details/2024/BlackRock-Launches-Its-First-Tokenized-Fund-BUIDL-on-the-Ethereum-Network-03-20-2024/default.aspx}.}, a tokenized institutional liquidity fund;
and 
Franklin Templeton's OnChain U.S. Government Money Fund, whose shares are represented by \texttt{BENJI} tokens\footnote{See \url{https://www.franklintempleton.com/press-releases/news-room/2026/franklin-templeton-stellar-development-foundation-mark-five-years-of-benji-the-first-u.s.-registered-tokenized-money-market-fund}.}. 
These products show that RWA tokenization is expanding beyond illiquid assets to conventional liquid financial instruments
and cash-management products.
\end{itemize}
As noted in \cite{Cong25},
tokenizing public equities and other already-liquid securities has a different economic motivation from tokenizing illiquid assets such as real estate or private credit.
Tokenization does not primarily create liquidity for an otherwise illiquid asset; instead, it can extend transferability beyond traditional exchange hours, facilitate integration with on-chain markets, and allow the security to interact with other blockchain-based financial contracts. 
Overall, RWA tokenization is viewed not merely as the digitization of an asset,
but as a redesign of the legal, trading, settlement, and information infrastructure of a financial claim.

\quad From financial perspective,
RWA tokenization is closely related to the theory of securitization,
which studies how cash flows from underlying assets are transformed into securities under asymmetric information.
For instance,
\cite{Le77} show how issuer retention can signal asset quality,
and 
\cite{Gorton90} study the creation of relatively information-insensitive claims to improve liquidity.
Particularly relevant to RWAs  is the DeMarzo-Duffie framework \cite{DD99},
where the security design tradeoffs between raising capital 
and the adverse-selection discount associated with private information.
Subsequent work \cite{De05} studies how pool and tranching can create liquid securities,
while \cite{Gorton07} considers the role of special-purpose vehicles in securitization.

\quad RWA tokenization inherits many of these classical issues. 
An issuer still chooses which cash flows to sell, how much exposure to retain, what legal claim investors receive, and how information about the underlying asset affects the price at issuance.
The main additional feature is that the resulting claim can be traded continuously on blockchain-based markets. 
This creates a dynamic dimension that is largely absent from traditional one-shot security-design models: 
after issuance, trading itself may reveal information about the underlying asset. 

\subsection{Stablecoins}
\label{sc22}

Stablecoins are digital tokens designed to maintain a stable value, typically one U.S. dollar. 
They differ primarily in the mechanisms used to maintain price stability, 
and can be broadly classified as fiat-backed, crypto-backed, and algorithmic stablecoins.

\quad The largest fiat-backed stablecoins are \texttt{USDT} and \texttt{USDC}.
\texttt{USDT}, issued by \texttt{Tether}, and \texttt{USDC}, issued by \texttt{Circle}, are backed by cash and short-term U.S. government securities.
Institutional customers can mint and redeem directly with the issuer, 
while the stablecoin itself trades permissionlessly across blockchains. 
This resembles a narrow bank or money-market fund,
where the issuer creates short-duration dollar liability and invests the corresponding reserves in safe and liquid assets. 
The temporary depeg of USDC following the 2023 collapse of Silicon Valley Bank illustrates their exposure to off-chain banking and reserve risks.

\quad Other stablecoins and tokenized money rely on different mechanisms.
\texttt{USDe}\footnote{See \url{https://docs.ethena.fi/} for details on \texttt{USDe}.}, developed by Ethena, 
is a synthetic dollar rather than a conventional fiat-backed stablecoin.
It combines crypto collateral with offsetting short positions in derivatives to hedge the dollar value of its collateral.
The recently introduced \texttt{Open USD}\footnote{See \url{https://www.onepay.com/newsroom/introducing-open-usd} for details on \texttt{Open USD}.} instead targets business payments, 
allowing participating firms to share reserve earnings and governance.
Traditional banks are also moving into tokenized money:
JPMorgan, for instance, developed \texttt{JPM Coin} and later \texttt{JPMD}, a tokenized commercial-bank deposit for institutional clients. 
Unlike the fiat-backed stablecoins, 
\texttt{JPMD} is a tokenized commercial-bank deposit, and hence remains a liability of JPMorgan rather than a reserve-backed token. 

\quad The canonical example of crypto-backed stablecoins is \texttt{DAI}, issued through MakerDAO, 
where users mint stablecoins against over-collateralized positions. 
The system uses liquidation, stability fees, and governance to maintain the peg.
Over time, however, DAI has increasingly incorporated fiat-backed and RWA collateral through Peg Stability Module,
illustrating that the boundary between crypto and conventional reserve backing is not strict.
At the other extreme are algorithmic stablecoins such as TerraUSD (\texttt{UST}), 
whose peg relied on endogenous arbitrage against another token rather than safe collateral. 
The collapse of \texttt{UST} in 2022 shows the fragility of such structures: when confidence in the supporting token deteriorated, redemption created additional supply of that token, generating a death spiral.

\quad The literature on stablecoins has developed around several related questions.
\cite{Cat22} studies stablecoin design and the tradeoff among stability, capital efficiency, and decentralization.
\cite{Lyons23} shows that arbitrage plays an important role in maintaining the peg (particularly for Tether),
while \cite{Ma26} points out a tradeoff between efficient arbitrage and run risk: 
easier redemption can stabilize prices but may also accelerate runs.
\cite{Gorton22} compares stablecoins with earlier forms of private money,
and studies  how stablecoin pegs can coexist with run risk.
\cite{Cas25, Li26} show that issuers may have incentive to invest reserves in higher-yielding but less liquid assets,
creating a classical maturity-transformation problem.
Overall, the central design problem for stablecoins is how reserve composition, redemption rules, arbitrage, and governance jointly sustain a stable peg. 

\subsection{DATs}
\label{sc33}

Digital Asset Treasuries (DATs) are publicly traded companies 
which hold a substantial amount of cryptoassets
and actively use capital markets to manage and expand these holdings. 

\quad The notable example is \texttt{Strategy} (formerly \texttt{MicroStrategy}),
which began accumulating Bitcoin in 2020,
and has subsequently transformed itself into a Bitcoin Treasury Company.
\texttt{Strategy} finances its Bitcoin accumulation through multiple capital-market instruments,
including common equity, convertible notes, and several classes of perpetual preferred stocks
(\texttt{STRF}, \texttt{STRC}, \texttt{STRK}, and \texttt{STRD}),
designed for investors with different yield, duration, conversion, and risk preferences.
A key feature of the Strategy model is the interaction between the market valuation of its equity and its ability to raise capital.
It uses common-stock at-the-market (ATM) offerings extensively when its market-to-NAV premium is high,
and uses preferred securities and convertible debt as alternatives. 
Recently, Strategy has explicitly described its capital structure in terms of {\em Digital Equity} and {\em Digital Credit},
reflecting an attempt to use different securities to attract different classes of capital.

\quad A second important example is \texttt{BitMine},
which adopts an Ethereum treasury strategy. 
Ethereum, however, is a Proof-of-Stake asset:
ETH can be staked to generate additional ETH, while participating in the validation of the Ethereum network. 
Thus, 
\texttt{BitMine} combines capital-market financing with protocol-level activities 
such as staking and potentially DeFi.
This provides a source of income that is absent for a Bitcoin DAT.
Indeed, \texttt{BitMine} emphasizes both strategic ETH accumulation and staking as central components of its treasury strategy. 
Similar example includes \texttt{Solana Company},
which is a Solana DAT that objective is to maximuizr sol per share
through both capital-market activity and Solana staking rewards from on-chain participation.
A similar example is \texttt{Solana Company},
a Solana DAT that seeks to maximize SOL per share through both capital-market activities 
and staking rewards.

\quad Other examples demonstrate further variations.
\texttt{Metaplanet} adopts a Bitcoin treasury strategy in Japan, 
and provides an example of the Strategy-type model outside the U.S.
\texttt{MARA Holdings}, \texttt{Riot} and \texttt{CleanSpark} also maintain substantial Bitcoin treasuries,
though their economics differ from \texttt{Strategy} 
because Bitcoin is generated partly through mining operations rather than acquired through capital-market financing. 
More recently, these mining companies have increasingly explored AI data centers as 
a diversification strategy alongside their mining businesses.
Ethereum treasury companies such as \texttt{SharpLink} emphasize staking income,
and hence, comparatively simpler balance sheets.

\quad The above examples show a key distinction between DATs and crypto ETFs.
A crypto ETF is primarily a passive investment vehicle, 
with creation and redemption mechanisms designed to keep its market price close to NAV.
A DAT, by contrast, has an endogenous capital structure: its equity may trade at a premium or discount to the NAV of its crypto holdings, and this valuation gap can affect its financing, and hence, crypto accumulation. 
This perspective explains why we include DATs in our broader taxonomy of RWA tokenization.
Standard RWA tokenization transforms an off-chain asset into a transferable on-chain claim. 
A DAT operates in the opposite direction: a DAT holds cryptoassets on its corporate balance sheet, and provides investors with exposure through traditional publicly traded securities.
In both cases,
the underlying asset is transformed into financial claims accessible to the investors.

\quad Compared with traditional securitization and stablecoins, 
the academic literature on DATs remains relatively sparse, reflecting the emergence of the industry. 
The core questions concern how a DAT chooses among alternative financing strategies for crypto accumulation, 
and how these choices interact with the company's market valuation 
and the characteristics of the underlying cryptoassets.

%-------------------------------------------------------------------------------------------------
\section{Model 1: Tokenization \& Securitization}
\label{sc5}
\quad In this section, we propose a simple model to illustrate how RWA tokenization relates to traditional securitization,
following the DeMarzo-Duffie framework \cite{DD99} on the role of private information in security design.

\quad As previously mentioned, securitization (or its pricing) is to create securities
whose payoffs are tied directly to underlying cash flows: 
\begin{equation*}
\mbox{Assets} \rightarrow \mbox{SPV (Special Purpose Vehicle)} \rightarrow \mbox{Claims}.
\end{equation*}
For on-chain RWAs, there is no essential difference at issuance:
\begin{equation*}
\mbox{Assets} \rightarrow \mbox{SPV/Protocol} \rightarrow \mbox{Tokens}.
\end{equation*}
Typical securitized assets, such as structured credit and private debt instruments,
trade relatively infrequently after initial issuance.
This is in contrast with RWAs, which are continuously traded on the secondary market/blockchains.
Thus, tokenization can turn a relatively static financing structure into a dynamically priced, continuously traded, and potentially continuously restructured security.
Our main theory is that tokenization complements securitization by adding a dynamic trading layer.
This trading layer narrows the adverse-selection discount, and enhances price discovery.
The structure of our model is as follows:
\begin{equation*}
\mbox{Issuer} \underset{\tiny \mbox{date } 0}{\overset{P_0}{\xrightarrow{\hspace*{1cm}}}}
\mbox{Investor} \stackrel{\tiny \mbox{secondary market}}{\xrightarrow{\hspace*{2cm}}} 
\mbox{Traders}
\end{equation*}

\subsection{Assets}
\label{sc51}
Time is continuous, indexed by $t \in [0, T]$, for a fixed $T>0$ representing the length of a finite horizon. 
We consider a simple model of RWA that delivers both a cash flow $(C_t)_{0 \le t \le T}$:
\begin{equation}
dC_t = \theta dt, \quad t \le T,
\end{equation}
and a terminal payoff:
\begin{equation}
X = x_0 + x_1 \theta + \varepsilon,
\end{equation}
where $\theta$ is private information available to the RWA issuer but unknown to the public;
$(x_0, x_1)$ is known to the public,
and $\varepsilon$ is a mean zero random variable. 
Here, $\theta$ can be rental earning of a real estate property;
repayment rate of a loan pool;
operating cash-flow rate of an infrastructure asset.
We also assume the risk-free rate $r > 0$.

\subsection{Security/tokenization design}
\label{sc52}
The problem of interest is to find a good (or optimal) design
$F((C_t)_{0 \le t \le T}, X)$.
For instance, for securitized assets with only a terminal payoff $X$,
$F_J(X) = (X - K)^+$ is the junior tranche,
and $F_S(X) = \min(X, K)$ is the senior tranche.

\quad Here we consider a special family of tokenization designs $F = (\alpha_C, \alpha_X)$, 
where $\alpha_C$ portion of the cash flow and $\alpha_X$ portion of the terminal payoff are tokenized. 
Denote by
\begin{equation}
dC^F_t = \alpha_C \theta dt, \quad X^F = \alpha_X X,
\end{equation}
for the tokenized assets.
Let’s explain a bit why not tokenize the entire assets.
Asymmetric information makes some RWA tokens hard to sell.
Investors on the public market observe the security design $F$, but not private information.
Then the issuance price reflects investors’ beliefs about the RWAs that issuers choose to sell.
A highly information-sensitive token (e.g., for large $\alpha_C, \alpha_X$) 
may therefore suﬀer a larger adverse-selection discount,
which is the core mechanism behind DeMarzo-Duffie's liquidity-based design.

\subsection{Private and public values}
Since the RWA token issuer knows the private information $\theta$, 
her value is:
\begin{equation}
\begin{aligned}
V^{\tiny \mbox{pvt}}_t &= \mathbb{E}\left[\int_t^T e^{-r(s-t)} dD_s^F + e^{-r(T-t)} X^F \,\Big|\, \theta\right] \\
& = \alpha_X e^{-r(T-t)} x_0 + H_t \theta,
\end{aligned}
\end{equation}
where 
\begin{equation}
\label{eq:H}
H_t:= \alpha_C \frac{1 - e^{-r(T-t)}}{r} + \alpha_X x_1 e^{-r(T-t)},
\end{equation}
is the token's sensitivity to private information (i.e., $H_t = \partial_\theta V^{\tiny \mbox{pvt}}_t$).
Note that $H_t$ depends on the tokenization design $F = (\alpha_C, \alpha_X)$.

\quad Now let's take the viewpoint from an uninformed trader in the secondary market.
Recall that the uninformed trader does not know the private information $\theta$,
but can observe some public trading information.
Here, assume that the uninformed trader observes the order flow $(Y_t)_{0\le t \le T}$,
and denote by $(\mathcal{F}_t)_{0 \le t \le T}$ its filtration.
Then her value (or market value) is:
\begin{equation}
\begin{aligned}
V^{\tiny \mbox{pub}}_t &= \mathbb{E}[V^{\tiny \mbox{pvt}}_t \,|\, \mathcal{F}_t] \\
& = \alpha_X e^{-r(T-t)} x_0 + H_t \, \mathbb{E}[\theta \,|\, \mathcal{F}_t].
\end{aligned}
\end{equation}
As a result, the information wedge (i.e., the difference between the informed and the market valuations) is:
\begin{equation}
\label{eq:delta}
\Delta_t = V^{\tiny \mbox{pvt}}_t - V^{\tiny \mbox{pub}}_t 
= H_t (\theta - \mathbb{E}[\theta \,|\, \mathcal{F}_t]).
\end{equation}

\subsection{Informed trading}
\label{sc53}
Assume that the uninformed trader observes the aggregate order flow:
\begin{equation}
\label{eq:Y}
dY_t = q_t dt + \sigma dB_t,
\end{equation}
where $q_t$ is the informed trader's trading rate (measured in token units per unit time),
and
$\sigma dB_t$ represents the noise trading ($\sigma > 0$ and $(B_t)_{0 \le t \le T}$ is Brownian motion).

\quad For an informed trader, she knows the private information $\theta$, 
and hence, has the information advantage $\Delta_t$. 
Assume that the price-impact cost is quadratic in the trading rate (i.e., Kyle's model \cite{Kyle85}).
The trader's objective is:
\begin{equation}
\max_{q_t} \left\{q_t \Delta_t - \frac{\gamma}{2} q_t^2\right\}, \quad \mbox{for some } \gamma > 0,
\end{equation}
which yields
\begin{equation}
\label{eq:q}
q^*_t = \frac{\Delta_t}{\gamma}.
\end{equation}
The associated profit rate is
$\Pi^*_t = \frac{\Delta_t^2}{2 \gamma}$.

\subsection{Information recovery}
\label{sc54}
Injecting \eqref{eq:q} into \eqref{eq:Y} yields:
\begin{equation}
dY_t = \frac{H_t (\theta - \mathbb{E}[\theta \,|\, \mathcal{F}_t])}{\gamma} dt + \sigma dB_t.
\end{equation}
Let 
\begin{equation}
v_t:= \var(\theta\,|\, \mathcal{F}_t).
\end{equation}
By Gaussian filtering, we obtain:
\begin{equation}
\frac{dv_t}{dt} = - \frac{H_t^2}{\gamma^2 \sigma^2} v_t^2,
\end{equation}
which gives:
\begin{equation}
\label{eq:vt}
v_t = \left(\frac{1}{v_0} + \frac{1}{\gamma^2 \sigma^2} \int_0^t H_s^2 ds\right)^{-1}.
\end{equation}
We see that $v_t$ decreases in $t$, 
so continuous trading generates information recovery.

\subsection{Issuance price}
The expected adverse-selection loss by uninformed traders is:
\begin{equation}
\mathbb{E}(\Pi^*_t \,|\, \mathcal{F}_t) =  \frac{H_t^2 v_t}{2 \gamma},
\end{equation}
and the total liquidity cost is:
\begin{equation}
\label{eq:Cliq0}
C^{\tiny \mbox{liq}} = \int_0^T e^{-\rho t} \frac{H_t^2 v_t}{2 \gamma}dt,
\end{equation}
where $\rho > 0$ is the discount of the secondary market.
Also note that $C^{\tiny \mbox{liq}}$ depends on the design $F = (\alpha_C, \alpha_X)$ through $H_t$.
Here, the term $H_t$ represents the tokenization design;
$v_t$ records the remaining private information;
and $\frac{1}{\gamma}$ underlies market liquidity and blockchain technology.

\quad Assume that date $0$ (issuance) investor anticipates future secondary market adverse-selection loss,
and expected costs are capitalized one-for-one into the issuance price.
We obtain the issuance price:
\begin{equation}
\label{eq:P0}
\begin{aligned}
P_0 &= V_0^{\tiny \mbox{pub}} - C^{\tiny \mbox{liq}} \\
& = \alpha_X e^{-rT} x_0 + H_0 \, \mu - 
 \int_0^T \frac{e^{-\rho t}  H_t^2}{2 \gamma} \left(\frac{1}{v_0} + \frac{1}{\gamma^2 \sigma^2} \int_0^t H_s^2 ds\right)^{-1}dt \\
 & = \frac{\alpha_C(1-e^{-rT})}{r}\mu+ \alpha_X e^{-rT}(x_0 + x_1 \mu) -  
 \int_0^T \frac{e^{-\rho t}  H_t^2}{2 \gamma} \left(\frac{1}{v_0} + \frac{1}{\gamma^2 \sigma^2} \int_0^t H_s^2 ds\right)^{-1}dt 
\end{aligned}
\end{equation}
where we denote $\mu: = \mathbb{E}[\theta]$.
Here, $V_0^{\tiny \mbox{pub}}$ is the ex-ante valuation of the tokenized assets $(C_t^F, X^F)$.

\subsection{Optimal tokenization design}
Assume that the issuer has valuable alternative investment opportunities,
so raising one dollar of capital creates incremental value $1 + \beta >1$ 
(here, $\beta$ is the retention cost).
The issuer's objective is:
\begin{equation}
\max_{F = (\alpha_C, \alpha_X)}\left\{(1+\beta) P_0 +  \frac{(1-\alpha_C)(1-e^{-rT})}{r}\mu + (1-\alpha_X) e^{-rT}(x_0 + x_1 \mu)\right\}.
\end{equation}
Equivalently, the objective is:
\begin{equation}
\label{eq:opt1}
\begin{aligned}
\max_{F = (\alpha_C, \alpha_X)}J(\alpha_C, \alpha_X) &= \beta V_0^{\tiny \mbox{pub}} - (1+\beta)C^{\tiny \mbox{liq}} \\
& = \beta \left( \frac{\alpha_C(1-e^{-rT})}{r} \mu + \alpha_X e^{-rT}(x_0 + x_1\mu) \right) \\
& \qquad \qquad -  
(1+\beta) \int_0^T \frac{e^{-\rho t}  H_t^2}{2 \gamma} \left(\frac{1}{v_0} + \frac{1}{\gamma^2 \sigma^2} \int_0^t H_s^2 ds\right)^{-1}dt,
\end{aligned}
\end{equation}
where $H_t$ is defined by \eqref{eq:H}.

\quad Now let's solve the optimization problem \eqref{eq:opt1}.
Denote by 
\begin{equation}
\label{eq:cab}
\begin{aligned}
& c_C: = \frac{1-e^{-rT}}{r} \mu, \quad c_X:=e^{-rT}(x_0 + x_1 \mu), \quad c =\begin{pmatrix} c_C \\ c_X \end{pmatrix}; \\
& a_t: = \frac{1-e^{-r(T-t)}}{r}, \quad b_t:=x_1 e^{-r(T-t)}.
\end{aligned}
\end{equation}
We then have $V_0^{\tiny \mbox{pub}} = \alpha_C c_C + \alpha_X c_X$ and $H_t = \alpha_C a_t + \alpha_X b_t$.
Thus,
\begin{equation}
\label{eq:HM}
\int_0^T H_t^2 dt = \alpha^T M \alpha,
\end{equation}
where  $M = \begin{pmatrix} M_{CC} & M_{CX} \\ M_{CX} & M_{XX} \end{pmatrix}$ is information sensitivity with
$M_{XX}: = \frac{x_1^2}{2r}(1 - e^{-2rT})$,
$M_{CX}: = \frac{x_1}{r^2} \left(1 - e^{-rT} - \frac{1-e^{-2rT}}{2} \right)$,
and $M_{CC}:= \frac{1}{r^2}\left(T - \frac{2(1-e^{-rT})}{r} + \frac{1-e^{-2rT}}{2r} \right)$,
and $\alpha =\begin{pmatrix} \alpha_C \\ \alpha_X \end{pmatrix}$.

\quad We consider the special case $\rho = 0$ (no discount for the secondary market),
which gives a closed-form solution.
Observing that $\frac{d}{dt} \log(v_0/v_t) = \frac{H_t^2 v_t}{\gamma^2 \sigma^2}$, we deduce:
\begin{equation}
\label{eq:Cliq}
\begin{aligned}
C^{\tiny \mbox{liq}} & = \frac{\gamma \sigma^2}{2} \log\left( \frac{v_0}{v_T}\right) \\
& = \frac{\gamma \sigma^2}{2} \log\left( 1 + \frac{v_0}{\gamma^2 \sigma^2} \int_0^T H_t^2 dt \right) = \frac{\gamma \sigma^2}{2} \log\left( 1 + \frac{v_0}{\gamma^2 \sigma^2} \alpha^T M \alpha \right).
\end{aligned}
\end{equation}
Let $\kappa:= \frac{v_0}{\gamma^2 \sigma^2}$. 
The objective \eqref{eq:opt1} becomes:
\begin{equation}
J(\alpha) = \beta c^T \alpha - \frac{(1+\beta) \gamma \sigma^2}{2} \log(1 + \kappa \alpha^T M \alpha).
\end{equation}

\quad Let's first consider the interior local maxima. 
The first-order condition yields:
\begin{equation}
\label{eq:FOC}
\beta c = (1+ \beta) \frac{\gamma \sigma^2 \kappa}{1+ \kappa \alpha^T M \alpha} M \alpha.
\end{equation}
Assuming that $M$ is non-singular, we get:
\begin{equation}
\label{eq:ac}
\alpha_{\tiny \mbox{int}} = \lambda M^{-1}c, \quad \mbox{for some } \lambda > 0.
\end{equation}
Denoting by $S: = c^T M^{-1}c$, and injecting \eqref{eq:ac} into \eqref{eq:FOC} yields:
\begin{equation}
\beta \kappa S \lambda^2 - \frac{(1+ \beta) v_0}{\gamma} \lambda + \beta = 0.
\end{equation}
Further assuming that $(1+\beta)^2 v_0 \sigma^2 \ge 4 \beta^2S$,
the local maximum corresponds to the smaller root 
\begin{equation}
\label{eq:lambdamin}
\lambda_{-}: = \frac{(1+\beta) \gamma \sigma^2}{2 \beta S}\left(1 - \sqrt{1 - \frac{4 \beta^2S}{(1+\beta)^2 v_0 \sigma^2}} \right).
\end{equation}
Putting all together gives:
\begin{equation}
\label{eq:int}
\alpha_{\tiny \mbox{int}} = \lambda_{-} M^{-1}c,
\end{equation}
where $c, M$ and $\lambda_{-}$ are defined in \eqref{eq:cab}, \eqref{eq:HM} and \eqref{eq:lambdamin} respectively.
The formula suggests that the RWA issuer prefer designs with high fundamental value relative to information sensitivity.

\quad Of course, the interior local maximum $\alpha_{\tiny \mbox{int}}$ is valid if it retains in $(0,1)$. 
We also need to check the local maxima on the four edges:
\begin{equation*}
(\alpha_C, 0), \, (\alpha_C, 1), \, (0, \alpha_X), \, (1, \alpha_X), \quad \mbox{with } \alpha_C, \alpha_X \in (0,1),
\end{equation*}
and the four corners $(0,0)$, $(0,1)$, $(1,0)$ and $(1,1)$.
We summarize the result as follows.
\begin{proposition}
\label{prop:51}
Assume that $\rho = 0$. 
Call $\alpha_{\tiny \mbox{int}}$ the feasible interior local maximum if $\lambda_{-} M^{-1}c \in (0,1)^2$.
Also call $\alpha_{\tiny \mbox{edge}}^i$, $i = 1,2,3,4$ the feasible edge local maximum if 
they are well-defined (as the real solution to some quadratic equations) and retain in $(0,1)^2$. 
Define 
\begin{equation}
\mathcal{A}: = \{\alpha_{\tiny \mbox{int}}, \alpha_{\tiny \mbox{edge}}^1, \alpha_{\tiny \mbox{edge}}^2, \alpha_{\tiny \mbox{edge}}^3, \alpha_{\tiny \mbox{edge}}^4, (0,0), (0,1), (1,0), (1,1)\}.
\end{equation}
Then the optimal tokenization design to \eqref{eq:opt1} is:
\begin{equation}
\alpha^* = \argmax_{\alpha \in \mathcal{A}} J(\alpha).
\end{equation}
\end{proposition}

\quad In other words, we obtain a complete solution to the optimal tokenization problem \eqref{eq:opt1}:
it suffices to solve five quadratic equations, and compare nine objective values.
We leave the full details of the (degenerate) edge cases to interested readers.

\subsection{Further comments and directions}
Several remarks are in order.

\quad First, the novel ingredients compared to DeMarzo-Duffie's static model are in Sections \ref{sc53}--\ref{sc54},
where continuous trading on the secondary market enhances price recovery.
This can be seen from the expression of $v_t$ in \eqref{eq:vt}: 
if there is no trading, we have
\begin{equation}
v_t \equiv v_0 > \left(\frac{1}{v_0} + \frac{1}{\gamma^2 \sigma^2} \int_0^t H_s^2 ds\right)^{-1},
\end{equation}
so the liquidity cost $C^{\tiny \mbox{liq}}$ in \eqref{eq:Cliq0} is larger,
and the issuance price $P_0$ in \eqref{eq:P0} is smaller.

\quad Second, let's assume that the optimal tokenization design is attained in the interior, 
i.e., $\alpha^* = \alpha_{\tiny \mbox{int}}$,
and examine the expression \eqref{eq:int}.
The parameters of interest are $(\sigma, \gamma, \beta)$,
and the dependence of $\alpha_{\tiny \mbox{int}}$ on these parameters is through $\lambda_{-}$.
Here are the interpretations:
\begin{itemize}[itemsep = 3 pt]
\item
If public information becomes more precise $\sigma \downarrow$, the market learns faster,
so adverse selection decreases and $\alpha^* \uparrow$.
Thus, better oracle and disclosure technology enhances tokenization.
\item
If market liquidity $\gamma \uparrow$,
then adverse selection narrows and $\alpha^* \uparrow$.
Thus, more efficient trading infrastructure supports greater tokenization.
\item
If the issuer has better (outside) investment opportunity $\beta \uparrow$,
there is greater value of upfront financing and $\alpha^* \uparrow$.
So the issuer tends to tokenize more, and put the issuance proceeds to more productive use.
\end{itemize}

\quad Next, we make the assumption $\rho = 0$ so as to obtain an explicit characterization of 
the optimal tokenization design. 
With $\rho > 0$, the factor $e^{-\rho t}$ prevents the liquidity integral \eqref{eq:Cliq0} from reducing to 
the logarithm \eqref{eq:Cliq},
so the first-order conditions are integral equations rather than quadratic equations.
Nevertheless, these integral equations can be solved numerically, 
and the structural result in Proposition \ref{prop:51} still holds.
Also note that by setting $\rho = 0$, 
we isolate the learning mechanism while preserving $r>0$ in cash-flow valuation.
 
\quad Finally, the DeMarzo-Duffie model trades off retention of cash flows against the illiquidity generated by making the security
more sensitive to private information.
In our setting, the amount of asymmetric information itself evolves, 
and liquidity is time dependent.
This is particularly natural for RWAs.
Imagine that initially the market knows little about a tokenized real-estate portfolio.
As time passes it observes rental payments, defaults, collateral values, redemptions, transactions,
and oracle updates.
So tokenization potentially goes beyond simply creating a liquid security:
\begin{equation*}
\mbox{trading} \to \mbox{learning} \to \mbox{reduced information asymmetry} \to \mbox{improved liquidity}.
\end{equation*}
Such a feedback does not appear naturally in a one-shot static security-design model.

\quad To conclude this section, we list a few directions for future work. 
\begin{enumerate}[itemsep = 3 pt]
\item
{\em Asset model}. 
In Section \ref{sc51}, we assume constant cash flows and linear terminal payoff,
which is a simplified model to illustrate our theory. 
In practice, the asset structure can be more complex.
For instance, it is reasonable to consider stationary cash flows $(\theta_t)_{0 \le t \le T}$,
and more general terminal payoff $G((\theta_t)_{0 \le t \le T})$.
\item
{\em Tokenization design}. 
We choose in Section \ref{sc52} a family of two-dimensional tokenization design $F = (\alpha_C, \alpha_X)$,
so the optimal tokenization problem \label{eq:opt1} is parametric. 
A further step is to consider broader classes of $F$, leading to a non-parametric mechanism design problem.
\item
{\em Trading mechanism}.
In Section \ref{sc53}, we adopt the Kyle's model, which assumes linear price impact, and hence, quadratic cost. 
Other price-impact relations are observed in practice, see e.g., \cite{DB15, GS11, TL11}.
Note that many RWAs are traded on blockchains or DeFi platforms. 
It is also interesting to consider blockchain-specific or AMM price-impact models in the design.
\item
{\em Data calibration}. 
This section is concerned with modeling and theory of RWA tokenization via securitization.
Our formulae can calibrate trading mechanisms and tokenization designs using historical issuance and trading data.
These calibrated models can then support the development of advanced financial instruments for RWAs.
\item
{\em Liquid asset tokenization}. 
Our theory emphasizes continuous trading as a key feature of RWA tokenization,
facilitating information recovery and reducing adverse-selection discount associated with illiquid assets. 
However, a growing class of tokenized RWAs consists of assets that are already highly liquid.
In this case, tokenization extends trading beyond the operating hours of traditional exchanges, 
creating an information wedge when the market is closed. 
An interesting problem is to extend our framework to how token trading during market closures aggregates information,
and predicts the underlying asset's price when the traditional market reopens.
\end{enumerate}

%-------------------------------------------------------------------------------------------------
\section{Model 2: Digital Asset Treasuries}
\label{sc6}

\quad In this section, we propose a simple model to illustrate the economics of DATs and their connection to corporate finance. 

\quad At first glance, a DAT resembles a crypto ETF:
investors obtain exposure to an underlying crypto through a publicly traded security.
The main distinction is that a DAT is an operating corporate balance sheet rather than a passive investment vehicle. 
In particular, its stock may trade at a premium or discount to the net asset value (NAV) of its crypto holdings,
and this valuation gap affects the firm's financing/investment decisions.
When the stock trades at a premium to NAV,
issuing equity can be accretive: the DAT raises capital, purchases more cryptos, 
and increases crypto holdings per share. 
This creates a feedback loop:
\begin{equation*}
\begin{aligned}
\text{market valuation} \to &  \text{ capital-market financing}  \\
& \qquad \to \text{crypto accumulation} \to \text{future market valuation}.
\end{aligned}
\end{equation*}
Our main theory is that this interaction between market valuation and financing distinguishes a DAT
from a passive crypto fund.
We develop a dynamic model that lets the DAT optimally chooses its issuance rate,
and the market-to-NAV premium is determined jointly with the firm's investment policy.

\subsection{DAT taxonomy}
\label{sc61}
We start by introducing important quantities to analyze DAT finance.
Time is continuous, indexed by $t \in [0, T]$, for a fixed $T>0$ representing the length of a finite horizon. 

\quad At time $t \in [0,T]$, 
let $P_t$ be the crypto price
and
$Q_t$ be the number of crypto units that the DAT holds.
For simplicity, we assume that the DAT only issues common stock to back crypto purchases.
Let $S_t$ be the DAT's common stock price in the secondary market,
and $N_t$ be its outstanding shares at time $t$.
Let
\begin{equation}
\label{eq:V}
q_t: = \frac{Q_t}{N_t} \quad \mbox{and} \quad V_t: = q_t P_t,
\end{equation}
be the number of crypto held per share,
and the NAV per share, respectively.

\quad The key to the analysis of the DAT finance is the market-to-NAV ratio:
\begin{equation}
\label{eq:m}
m_t: = \frac{S_t}{V_t}.
\end{equation}
This quantity distinguishes DATs from crypto ETFs.
For crypto ETFs, the creation and redemption mechanism enforces 
$m_t^{\tiny \mbox{ETF}} \approx 1$, i.e., crypto ETFs are NAV-neutral.
This is not necessarily the case for DATs:
$m_t^{\tiny \mbox{DAT}} > 1$ implies that common issuance is accretive;
$m_t^{\tiny \mbox{DAT}} < 1$ means that common issuance is dilutive. 
Typically, DATs want $m_t^{\tiny \mbox{DAT}} > 1$ so that they can issue more common shares for crypto accumulation.

\subsection{Common share issuance}
\label{sc62}
Assume that the DAT issues common shares at proportional rate $a_t > 0$: 
\begin{equation}
dN_t = a_t N_t dt.
\end{equation}
Let all issuance proceeds be used to purchase crypto, 
with a quadratic cost rate $\frac{\gamma}{2} a_t^2 V_t N_t$. 
Here, $\gamma>0$ represents balance-sheet adjustment frictions.
The specification scales this cost by the DAT's NAV $V_tN_t$, preserving homogeneity of the model.
So the balance sheet is:
\begin{equation}
\label{eq:P}
P_t dQ_t = S_t dN_t - \frac{\gamma}{2} a_t^2 V_t N_t dt,
\end{equation}
which gives the DAT accretion equation:
\begin{equation}
\label{eq:q}
dq_t  = \left((m_t - 1)a_t - \frac{\gamma}{2} a_t^2\right) q_t dt.
\end{equation}

\subsection{NAV dynamics}
\label{sc63}
Assume that the crypto price $P_t$ is governed by geometric Brownian motion:
\begin{equation}
dP_t = \mu P_t dt + \sigma P_t dB_t,
\end{equation}
where $\mu$ is the crypto growth rate, $\sigma > 0$ is the price volatility, and $(B_t)_{0 \le t \le T}$ is standard Brownian motion.
Combining \eqref{eq:V}, \eqref{eq:P} and \eqref{eq:q} yields the dynamics of NAV:
\begin{equation}
\label{eq:Veq}
dV_t = \left( \mu + (m_t - 1)a_t - \frac{\gamma}{2} a_t^2\right) V_t dt + \sigma V_t dB_t.
\end{equation}
In the drift term of \eqref{eq:Veq},
$\mu$ is the crypto growth; 
$(m_t-1)a_t$ represents the capital market accretion;
and $-\frac{\gamma}{2} a_t^2$ is an issuance or adjustment cost.
In dollar terms, this cost is proportional to the DAT's NAV, $\frac{\gamma}{2}a_t^2V_tN_t$.
Thus, $\gamma$ captures the increasing marginal frictions associated with financing the balance sheet, 
rather than the price impact of purchasing crypto in the secondary market.
We also assume the risk-free rate $r>0$.

\subsection{Optimal issuance rate}
\label{sc64}
Here, we formulate the optimal issuance  as an equilibrium control problem. 
First, we assume that the DAT's common share price depends on its NAV:
\begin{equation}
S_t = F(t, V_t),
\end{equation}
where $F: [0,T] \times \mathbb{R}_+ \to \mathbb{R}_+$ is a priori unknown
but is increasing in $v$.
By \eqref{eq:m}, $m_t = F(t, V_t)/V_t$, 
and injecting it into \eqref{eq:Veq} yields:
\begin{equation}
dV_t = \left(\mu V_t +a_t(F(t,V_t) - V_t) -\frac{\gamma}{2}a_t^2 V_t \right) dt + \sigma V_t dB_t.
\end{equation}

\quad The DAT management aims to maximize the value of the incumbent share:
\begin{equation}
J(t, v) = \max_{a_t > 0} \mathbb{E}\left[e^{-r(T-t)} V_T \, \bigg| \, V_t = v \right].
\end{equation}
but the price at which DAT can issue shares is itself $S_t = F(t,V_t)$.
This leads to the rational price equilibrium:
\begin{equation}
\label{eq:DP}
S_t = J(t, V_t) \Longleftrightarrow F(t, V_t) = \max_{a_t} \mathbb{E}\left[e^{-r(T-t)} V_T \, \bigg| \, V_t = v \right].
\end{equation}
Applying dynamic programing to \eqref{eq:DP},
we deduce the Hamilton-Jacobi-Bellman equation:
\begin{equation}
\label{eq:HJB}
\left\{ \begin{array}{lcl}
\partial_t F + \mu v \partial_v F +\frac{1}{2} \sigma^2 v^2 \partial_{vv} F - r F + \sup_{a > 0} \left[ \left(a(F - v) - \frac{\gamma}{2} a^2 v\right) \partial_v F\right] = 0, \\ 
F(T,v) = v.
\end{array}\right.
\end{equation}

\quad Next, let's solve the equation \eqref{eq:HJB}. 
The supremum problem in \eqref{eq:HJB} gives:
\begin{equation}
\label{eq:astar}
a^*(t,v) = \frac{(F(t,v) - v)_{+}}{\gamma v} = \frac{(m_t - 1)_{+}}{\gamma},
\end{equation}
which implies that a larger premium generates more issuance.
Substituting \eqref{eq:astar} back to \eqref{eq:HJB} yields:
\begin{equation}
\label{eq:HJ}
\partial_t F + \left(\mu v + \frac{(F - v)_+^2}{2 \gamma v} \right) \partial_v F +\frac{1}{2} \sigma^2 v^2 \partial_{vv} F - r F = 0.
\end{equation}
By homogeneity in $v$, the solution to the equation takes the separable form:
\begin{equation}
\label{eq:separable}
F(t,v) = f(t)v,
\end{equation}
so the market-to-NAV ratio $m_t = f(t)$.
Injecting \eqref{eq:separable} into \eqref{eq:HJ} yields the ordinary differential equation (ODE):
\begin{equation}
\label{eq:ODE}
f'(t) + (\mu - r)f(t) + \frac{f(t)(f(t)-1)_{+}^2}{2 \gamma} = 0, \quad f(T) = 1.
\end{equation}

\quad The solution to \eqref{eq:ODE} depends on the sign of $\mu -r$. 
We summarize the result in the following proposition.
\begin{proposition}
\label{prop:61}
The maximal solution to \eqref{eq:ODE} is given as follows.
\begin{enumerate}[itemsep = 3 pt]
\item
For $\mu < r$, we have $f(t) = e^{(\mu - r)(T-t)}$ for $0 \le t \le T$.
\item
For $\mu = r$, we have $f(t) \equiv 1$ for $0 \le t \le T$.
\item
For $\mu > r$, define
\begin{equation}
\eta:= 2 \gamma (\mu - r) \quad \mbox{and} \quad \tau_*:=\frac{\gamma}{1+\eta} \left(\log \eta + \frac{\pi}{\sqrt{\eta}} \right).
\end{equation}
The solution exists only for $0 \le T-t < \tau_*$, and is uniquely characterized by
\begin{equation}
\label{eq:ft}
T-t = \frac{2 \gamma}{1+ \eta}\left( \log f(t) - \frac{1}{2} \log \left( \frac{(f(t)-1)^2 + \eta}{\eta} \right) + \frac{1}{\sqrt{\eta}} \arctan\left( \frac{f(t)-1}{\sqrt{\eta}}\right) \right).
\end{equation}
Moreover, the solution exists on $[0,T]$ if and only if $T < \tau_*$.
\end{enumerate}
\end{proposition}

\quad The proof of the proposition is fairly standard, and is left to the readers.
It identifies three regimes for DAT's financial management:
\begin{itemize}[itemsep = 3 pt]
\item
If $\mu < r$, then $m_t = e^{(\mu - r)(T-t)} < 1$ and $a_t^* = 0$.
\item
If $\mu = r$, then $m_t \equiv 1$ and $a_t^* = 0$.
\item
If $\mu > r$, then $m_t = f(t) > 1$ given by \eqref{eq:ft} and $a_t^* = \frac{m_t - 1}{\gamma}$.
\end{itemize}
Thus, financing becomes active only when the crypto asset has sufficiently high expected growth ($\mu>r$).
This generates a premium to NAV ($m_t>1$), 
which makes equity issuance accretive.
For $\mu > r$, 
the market-to-NAV ratio $m_t=f(t) \downarrow 1$ as $t \uparrow T$.
Here, $T$ can be interpreted as the effective horizon of the premium-to-NAV financing opportunity:
the DAT initially benefits from a valuation premium that makes common-equity issuance accretive, 
but this financing advantage gradually dissipates.
When $m_T=1$, the optimal common issuance falls to zero.
Also note that $\tau_*$ is decreasing in $\eta$, and $\lim_{\eta \downarrow 0} \tau_*(\eta)  = + \infty$.
So the condition $T < \tau_*$ is satisfied for sufficiently small $\eta$. 

\subsection{Alternative financial instruments}
\label{sc65}
So far we have assumed that DAT finances crypto purchases only through common equity. 
DATs can also issue preferred stock, debt and convertible securities. 
Here, we give a simple extension of the model in Sections \ref{sc62}--\ref{sc63}.
that allows us to unify these financing instruments.

\quad Recall from \eqref{eq:Veq} that the NAV growth generated by common equity issuance is:
\begin{equation}
(m_t-1)a_t-\frac{\gamma_{\tiny \mbox{com}}}{2}a_t^2,
\qquad \mbox{with } \gamma_{\tiny \mbox{com}}=\gamma.
\end{equation}
In other words, 
the marginal financing advantage of common equity is $h_{\tiny \mbox{com}}(m):=m-1$.
More generally, 
let $a_{j,t}\geq 0$ be the issuance rate of security $j$,
and let its contribution to common-share NAV be
\begin{equation}
\label{eq:hgagen}
h_j a_{j, t} - \frac{\gamma_j}{2} a_{j,t}^2,
\end{equation}
where $h_j$ is the net financing advantage, and $\gamma_j>0$ captures issuance frictions and market capacity.
Besides common equity, we have:
\begin{itemize}[itemsep = 3 pt]
\item
Preferred stock:
It pays a fixed dividend, and sits above common stock but below debt in priority during a liquidation. 
For instance, \texttt{Strategy} issues preferred stocks \texttt{STRC}, \texttt{STRD}, \texttt{STRK}, and \texttt{STRF}.
Adopting the form \eqref{eq:hgagen}, the optimal preferred stock issuance is 
$a^*_{\tiny \mbox{pre},t} = \frac{(h_{\tiny \mbox{pre}})_+}{\gamma_{\tiny \mbox{pre}}}$,
which yields the NAV growth $\frac{(h_{\tiny \mbox{pre}})^2_+}{2 \gamma_{\tiny \mbox{pre}}}$.
Here, $h_{\tiny \mbox{pre}}$ is (almost) independent of $m$.
\item
Debt: Similar to preferred stock, the optimal issuance is 
$a^*_{\tiny \mbox{deb},t} = \frac{(h_{\tiny \mbox{deb}})_+}{\gamma_{\tiny \mbox{deb}}}$,
which yields the NAV growth $\frac{(h_{\tiny \mbox{deb}})^2_+}{2 \gamma_{\tiny \mbox{deb}}}$.
Also $h_{\tiny \mbox{deb}}$ is independent of $m$.
\item
Convertible notes: It is debt financing but grants the investor an equity conversion option.
Thus, its financing advantage $h_{\tiny \mbox{conv}}(m)$ is decreasing in $m$.
\end{itemize}
When the stock is cheap, granting investors an equity conversion option is relatively inexpensive. 
But when the stock already trades at a very high premium, 
issuing common stock directly becomes very attractive.

\quad Now assume for simplicity that DAT chooses one financing instrument at a time.
Define
\begin{equation}
\Phi(m):= \max\left\{ \frac{(m-1)_{+}^2}{2 \gamma_{\tiny \mbox{com}}}, \frac{(h_{\tiny \mbox{pre}})^2_+}{2 \gamma_{\tiny \mbox{pre}}},
\frac{(h_{\tiny \mbox{deb}})^2_+}{2 \gamma_{\tiny \mbox{deb}}},
\frac{(h_{\tiny \mbox{conv}}(m))^2_+}{2 \gamma_{\tiny \mbox{conv}}} \right\}.
\end{equation}
The ODE \eqref{eq:ODE} changes to:
\begin{equation}
f'(t) + (\mu - r)f(t) + f(t) \Phi(f(t)) = 0, \quad f(T) = 1.
\end{equation}

\subsection{Case study: Strategy}
\label{sc66}

To demonstrate the application of our theory, 
we give a ``back-of-envelop" analysis of the DAT \texttt{Strategy}, 
which initiated Bitcoin accumulation in August, 2020.
Table \ref{tab:MSTR} presents \texttt{Strategy}'s quarterly Bitcoin holdings, common stock prices, and market-to-NAV ratios from December 2024 to June 2026.\footnote{The data of Strategy's Bitcoin holdings and common shares outstanding is from Form 10-Q available at \url{https://www.strategy.com/financial-documents}.}

\begin{table}[ht]
\centering
\caption{Strategy: BTC holdings and market-to-NAV ratios.}
\label{tab:MSTR}
\begin{tabular}{cccccc}
\hline
Date 
& BTC holdings $Q_t$
& Shares $N_t$ (m)
& BTC price $P_t$
& MSTR price $S_t$
& $m_t$ \\
\hline
Dec.\ 31, 2024 & 447,470 & 245.778 & \$93,390  & \$289.62 & 1.70 \\
Mar.\ 31, 2025 & 528,185 & 266.177 & \$82,445  & \$288.27 & 1.76 \\
Jun.\ 30, 2025 & 597,325 & 280.958 & \$107,752 & \$404.23 & 1.77 \\
Sep.\ 30, 2025 & 640,031 & 287.109 & \$114,378 & \$322.21 & 1.26 \\
Dec.\ 31, 2025 & 672,500 & 312.062 & \$87,515  & \$151.95 & 0.81 \\
Mar.\ 31, 2026 & 762,099 & 346.222 & \$67,773  & \$124.80 & 0.84 \\
Jun.\ 30, 2026 & 846,000 & 371.603 & \$58,714  & \$86.93  & 0.65 \\
\hline
\end{tabular}
\end{table}
\quad During this period, \texttt{Strategy}'s market-to-NAV ratio decreased from a premium of $1.7$ to a discount of $0.65$. 
Of particular interest is the year 2025, when the ratio dropped from above 1.0 to below 1.0.
Table \ref{tab:MSTR25} records \texttt{Strategy}'s quarterly common shares issued\footnote{Here, the common shares issuance (for Bitcoin accumulation) is not exactly the difference between the shares outstandings in Table \ref{tab:MSTR}.
It excludes employee equity, conversions/exercises, Class B changes, etc.}
and annualized issuance intensities (computed by $a_t/\Delta t$, $\Delta t = \frac{1}{4}$).
\begin{table}[ht]
\centering
\caption{Strategy's common equity issuance and market-to-NAV ratios}
\label{tab:MSTR25}
\begin{tabular}{cccc}
\hline
Quarter & Class A common shares issued & Annualized issuance rate & $m_t$  \\
\hline
2025Q1 & 12.625m & 0.205 & 1.76 \\
2025Q2 & 14.226m &0.213 & 1.77 \\
2025Q3 &  5.712m & 0.081 & 1.26 \\
2025Q4 & 24.769m & 0.345 & 0.81 \\
\hline
\end{tabular}
\end{table}

\quad The key to our analysis is Proposition \ref{prop:61} that characterizes the market-to-NAV ration $m_t$,
and the optimal issuance formula \eqref{eq:astar} that gives $a_t^*= (m_t-1)_{+}/\gamma$.
Consider the high premium quarters 2025Q1--2025Q3.
By estimating $\hat{\mu} = 0.27$ and $r = 0.043$,
we get:
\begin{equation}
\hat{\delta} = 0.227.
\end{equation}
Then we use the formula \eqref{eq:ft} to calibrate $\gamma$.
Table \ref{tab:gamma} gives the least-square estimates of $\gamma$ for different time horizons $T$.
\begin{table}[htbp]
\centering
\caption{Estimated $\gamma$ for different $T$'s.}
\label{tab:gamma}
\begin{tabular}{c|ccccccccc}
\hline
$T$ (years)
& 1.00 & 1.25 & 1.50 & 1.75 & 2.00 & 2.25 & 2.50 & 2.75  \\
\hline
$\widehat{\gamma}$
& 0.020 & 0.057 & 0.128 & 0.252 & 0.470 & 0.898 & 1.980 & 8.498  \\
\hline
\end{tabular}
\end{table}
On the other hand, we can also use the formula \eqref{eq:astar} to calibrate $\gamma$,
which gives $\hat{\gamma}' \approx 3.62$.
Matching the two calibrated $\gamma$ yields:
\begin{equation}
T = 2.63 \mbox{ years}.
\end{equation}
Thus, if \texttt{Strategy} planned to finance its crypto purchases over a horizon longer than $2.63$ years, 
the model suggests that it over-issued common stock during this period. 
Conversely, if the planned financing horizon was shorter than $2.63$ years,
the model suggests that it under-issued common stock.

\quad In 2025Q4, however, the market-to-NAV ratio dropped below 1.0,
so the Bitcoin-accretive common issuance implies $a^* = 0$,
which seems to contradict the observation $a^* \approx - 0.345$.
In fact, \texttt{Strategy} used common-equity proceeds not only to acquire Bitcoin,
but also to fund preferred dividends, convertible-note interest, and later its USD reserve.
As noted by the company, ``The issuance of class A common stock increased Assumed Diluted Shares Outstanding without a corresponding increase to our bitcoin holdings".\footnote{This is excerpted from \texttt{Strategy}'s SEC filing \url{https://www.sec.gov/Archives/edgar/data/1050446/000105044626000020/mstr-20251231.htm}.}

\quad We also point out that in 2025Q1, \texttt{Strategy} raised about $\$4.4$ billion from common equity issuance along with
$\$1.98$ billion from convertible notes and $\$1.3$ billion from preferred financing.
But in 2025Q3, common equity proceeds fell to about $\$2.2$ billion while preferred financing rose to about $\$ 3.9$ billion.
This is consistent with our claim in Section \ref{sc65} that 
when the market-to NAV ratio $m$ decreases, issuing common equity becomes less attractive relative to debt/preferred financing.

\subsection{Further comments and directions}
Now let's make a few remarks, and mention several directions for future research.

\quad First, in Sections \ref{sc62}--\ref{sc63}, we assume a quadratic issuance cost and that crypto prices follow geometric Brownian motion. 
These assumptions are made for tractability, 
allowing us to obtain explicit solutions and economic interpretations.
The modeling choices are deliberately parsimonious and may not fully capture market practice.
In particular, the quadratic term in \eqref{eq:P}
should be viewed as a balance-sheet adjustment cost:
it is proportional to the DAT's NAV, rather than being a direct model of the price impact generated by its crypto purchases.
A richer model could distinguish these financing frictions from trading costs in the crypto market.
Though a substantial fraction of DAT's crypto purchases may be executed through OTC markets, 
large DATs may use OTC and other execution mechanisms to mitigate the immediate price impact of crypto purchases,
their large and persistent demand may nevertheless affect equilibrium crypto prices.
Investment decisions would then generate endogenous crypto-market price impact from the associated purchases.
Moreover, when multiple DATs hold and trade the same cryptoasset, 
their investment decisions may interact through aggregate market impact,
leading to a mean-field model (see e.g., \cite{TangMF26}).

\quad Second, our main theory in Section \ref{sc64} assumes that the DAT issues only common equity to finance crypto accumulation,
though Section \ref{sc65} briefly discusses extensions to other financing instruments
such as preferred equity, convertible notes, and debt.
The analysis, however, provides only a simplified treatment of these instruments.
An important future direction is to incorporate senior securities into a unified dynamic security-design problem.
For instance, we can consider a commitment framework as in \cite{DeHe16},
in which shareholders dynamically choose both common and senior financing.
Such an extension would naturally lead to a two-dimensional dynamic optimization problem, 
with leverage and the NAV premium as state variables.

\quad Next, in the case study in Section \ref{sc66}, 
we estimate the crypto growth rate $\mu$ and  use it to infer the duration of the premium-to-NAV financing opportunity.
This allows us to assess whether \texttt{Strategy} over- or under-issued common equity to finance its crypto accumulation.
An alternative approach is to take the financing horizon $T$ as given--
for instance, by assuming that the DAT plans a multi-year common-equity issuance program to finance crypto purchases,
and use the same framework to infer the crypto growth rate that matches the the observed issuance rate. 
Moreover, since \texttt{Strategy} conducted frequent, often weekly, equity issuances and crypto purchases during 2024--2026, 
a natural extension is to exploit higher-frequency data to estimate the model statistically,
and test whether the company's observed issuance decisions are consistent with the model-implied financing rule.

\quad Finally, it would be interesting to study DATs holding different types of cryptoassets. 
Our analysis, motivated primarily by \texttt{Strategy}, 
focuses on a Bitcoin DAT, where value creation arises mainly from capital-market financing, 
accretive crypto accumulation, and crypto price appreciation. 
Other DATs may have fundamentally different mechanisms. 
For example, \texttt{BitMine} can earn staking rewards by deploying its ETH in Ethereum validation, so its NAV depends not only on financing, ETH accumulation, and price appreciation, but also on staking income. 
Moreover, validator participation may create feedback between the DAT's investment policy and the underlying blockchain ecosystem. 
This suggests a broader theory of heterogeneous DATs, where optimal financing and accumulation depend on the characteristics of the underlying cryptoasset.
In particular, extending the model to incorporate staking yields, endogenous validator participation, and feedback between DAT holdings and blockchain fundamentals may help explain why the optimal financing strategy of a Proof of Stake-based DAT 
differs from that of a Bitcoin DAT.

%-------------------------------------------------------------------------------------------------
\section{Conclusion}
\label{sc4}

\quad This paper provides a gentle introduction to RWA tokenization,
and develops simple economic models to understand some of its key mechanisms. 
We propose to classify RWA tokenization into three broad categories:
standard RWA tokenization, stablecoins, and Digital Asset Treasuries (DATs), 
which are naturally connected to securitization, monetary economics, and corporate finance, respectively.
Our first model extends classical security design by incorporating continuous secondary-market trading, through which information is gradually revealed and the adverse-selection cost declines. 
Our second model studies DATs, where market valuation, capital-market financing, and crypto accumulation are jointly determined.

\quad The models are deliberately parsimonious, and leave several directions for future research. 
For RWA tokenization, 
these include richer cash-flow structures, broader security design, and blockchain-specific trading mechanisms such as AMMs. 
For DATs,
important extensions include endogenous alternative financing instruments and heterogeneous cryptoassets with staking or other on-chain returns.
Both models also suggest empirical questions. 
Issuance and secondary-market data can be used to estimate information recovery and liquidity effects in tokenized assets, while higher-frequency DAT data can be used to estimate financing rules and test the relation between NAV premiums, security issuance, and crypto accumulation.

\quad Looking ahead, RWA tokenization may also become part of the infrastructure of an agentic AI economy. 
AI agents may need to make payments, hold and exchange assets, purchase services, and transact with other agents continuously, making programmable digital money and tokenized assets natural building blocks for machine-to-machine commerce. 
At the same time, resources required by AI itself, e.g., GPUs, energy, data, and potentially AI-generated intellectual property,
could be represented and traded through tokenized claims.
This idea is closely related to Decentralized Physical Infrastructure Networks (DePIN) \cite{Lin25, J26}, 
where blockchains are used to coordinate distributed physical resources such as computing, storage, communication networks, and energy infrastructure.
This suggests a broader vision in which blockchain provides programmable ownership and settlement, while AI agents provide autonomous economic decision-making.
Understanding the interaction between these two technologies may open a new direction for tokenization economics.

\bigskip
{\bf Acknowlegdment}: This research is supported by NSF CAREER Award DMS-2538791, 
NSF MFAI Award DMS-2602038 and 
a Columbia-CityU/HK
collaborative project that is supported by InnoHK Initiative, The Government of the HKSAR
and the AIFT Lab.

\bibliographystyle{abbrv}
\bibliography{unique}
\end{document}